\documentclass{iaus}
\usepackage{graphicx}

\title[Star Formation in Different Galaxy Types] 
{Observational Comparison of Star Formation \\ in Different Galaxy Types}

\author[E.K.\ Grebel]   
{Eva K. Grebel$^1$}

\affiliation{$^1$Astronomisches Rechen-Institut, Zentrum f\"ur
Astronomie der Universit\"at Heidelberg,\\ M\"onchhofstr.\ 12--14, 
D-69120 Heidelberg, Germany\\ email: {\tt grebel@ari.uni-heidelberg.de}} 

\pubyear{2010}
\volume{270}  
\pagerange{1--12}
\jname{Computational Star Formation}
\editors{J. Alves, B.G. Elmegreen, J.M. Girart, \& V. Trimble, eds.}
\begin{document}

\maketitle

\begin{abstract} Galaxies cover a wide range of masses and star
formation histories.  In this review, I summarize some of the
evolutionary key features of common galaxy types.  At the high-mass
end, very rapid, efficient early star formation is observed,
accompanied by strong enrichment and later quiescence, well-described
by downsizing scenarios. In the intermediate-mass regime, early-type
galaxies may still show activity in low-mass environments or when
being rejuvenated by wet mergers.  In late-type galaxies, we find
continuous, though variable star formation over a Hubble time.  In the
dwarf regime, a wide range of properties from bursty activity to
quiescence is observed.  Generally, stochasticity dominates here, and
star formation rates and efficiencies tend to be low.  Morphological
types and their star formation properties correlate with environment.  
\keywords{galaxies: formation, galaxies: evolution, galaxies:
elliptical and lenticular, galaxies: spiral, galaxies: dwarf,
galaxies: stellar content}
\end{abstract}

\firstsection 
\section{Introduction}

Star formation differs widely in different galaxy types (e.g.,
\cite[Kennicutt 1998]{Kennicutt98}), ranging from slow, low-efficiency
events that may be long-lasting to intense, short-duration starbursts.
Along the Hubble sequence, typical global present-day star formation
rates range from $\sim 0$ M$_{\odot}$ yr$^{-1}$ in giant ellipticals
to $\sim 20$ M$_{\odot}$ yr$^{-1}$ in gas-rich, late-type spirals.
Starburst galaxies show star formation rates of up to $\sim 100$
M$_{\odot}$ yr$^{-1}$, and ultra-luminous infrared galaxies (ULIRGs)
appear to form up to $\sim 1000$ M$_{\odot}$ yr$^{-1}$ in stars.  Star
formation may be localized or encompass a large fraction of the
baryonic gas mass of a galaxy.  It may be continuous, declining or
increasing in intensity, or episodic.  It may be triggered by internal
processes within a galaxy or by interactions with other galaxies.
Typical sites of present-day star formation in galaxies are located in
the extended disks of spirals and irregulars, in the dense gas disks
in galaxy centers (circumnuclear star formation), and in regions of
compressed gas in starbursts, interacting galaxies, or tidal tails.
The star formation histories of galaxies vary with galaxy type, mass,
gas content, and environment as well as with time, and are tightly
coupled with their chemical evolution.    

\section{Very early star formation}

Giant elliptical galaxies are found to have experienced the bulk of
their star formation at early times and to have undergone enrichment
very rapidly.  A similar evolution appears to have taken place in the
bulges of spiral galaxies.  But how early is ``early''?  Age dating
resolved old stellar field populations in nearby galaxies remains
difficult due to crowding, extinction, the superposition of stellar
populations of different ages, and the general difficulty of
associating individual red giants with a specific age.  These problems
are exacerbated in galaxies where only integrated-light properties can
be analyzed, which makes it very difficult to date stellar populations
older than a few Gyr.

\subsection{Quasars at redshift $\ge 6$}

An alternative approach, limited to very luminous and hence presumably
very massive, actively star-forming objects, is the analysis of
galaxies observed at high redshift.  Surprisingly, even the very young
quasars discovered at a redshift of $z \sim 6$ (age of the Universe:
$\sim$ 900 Myr) reveal metal absorption lines that translate into
supersolar metallicities, indicating very rapid, early enrichment.
These objects may be the precursors of giant ellipticals (\cite[Fan et
al.\ 2001]{Fan01}).  The masses of the central black holes in  $z \sim
6$ QSOs are estimated to range from several $10^8$ to several $10^9$
M$_{\odot}$.  This suggests formation redshifts of more than 10 for
putative 100 M$_{\odot}$ seed black holes if continuous Eddington
accretion is assumed.  Other, more rapid black hole formation
mechanisms may also play a role (e.g., \cite[Fan 2007]{Fan07}).  

For one of the highest-redshift quasars known, SDSS
J114816.16+525150.3 at $z = 6.42$, sub-millimeter and radio
observations suggest a molecular gas mass (CO) of $\sim
5\times10^{10}$ M$_{\odot}$ within a radius of 2.5 kpc around the
central black hole if the gas is bound.  The star formation rate in
this object is $\sim 1000$ M$_{\odot}$ yr$^{-1}$, akin to what has
been derived for star-bursting ultraluminous infrared galaxies.  Even
within 20 kpc, the inferred mass estimate for the QSO's host is still
comparatively low, $\sim 10^{11}$ M$_{\odot}$.  A central
massive $10^{12}$ M$_{\odot}$ bulge has evidently not yet formed
(\cite[Walter et al.\ 2004] {Walter04}; \cite[Fan 2007]{Fan07}).  

In several $z\gtrsim 6$ QSOs dust has been detected.  One of the
primary sources for dust in the present-day Universe are low- and
intermediate-mass asymptotic giant branch stars, but these need at
least some 500 Myr to 1 Gyr to begin generating dust in their
envelopes.  In quasars in the early Universe, other mechanisms are
believed to be responsible, including dust production in supernovae of
type II (\cite[Maiolino et al.\ 2004]{Maiolino04}) and in quasar winds
(\cite[Elvis et al.\ 2002]{Elvis02}).  Recently two $z \sim 6$ QSOs
{\em without dust} were detected, which indicates that these objects
are probably first-generation QSOs forming in an essentially dust-free
environment (\cite[Jiang et al.\ 2010] {Jiang10}).  This discovery
illustrates that even at these very early times of less than 1 Gyr
after the Big Bang, galactic environments differ in the onset of
massive star formation and in the degree of heavy-element pollution,
with some regions already having experienced substantial enrichment.
One may speculate that the densest regions are the first ones to start
massive star formation, and that the dusty high-redshift QSOs trace
these regions. 

Magnesium, an $\alpha$ element, is produced in supernovae of type II
and hence expected to be generated soon after massive star formation
commences.  Iron is predominantly produced in supernovae of type Ia,
requiring a minimum delay time of $\sim 300$ Myr when instantaneous
starbursts are considered (e.g., \cite[Matteucci \& Recchi
2001]{Matteucci01}).  Fe is detected in $z \sim 6$ QSOs despite their
young age. The Fe\,{\sc ii}/Mg\,{\sc ii} ratio (which is essentially a
proxy for Fe/$\alpha$) is found to be comparable to that of
lower-redshift QSOs (e.g., \cite[Barth et al.\ 2003]{Barth03};
\cite[Freudling et al.\ 2003]{Freudling03}; \cite[Kurk et al.\
2007]{Kurk07}) and to be around solar or super-solar metallicity.  The
near-constancy of the Fe\,{\sc ii}/Mg\,{\sc ii} ratio as a function of
redshift implies a lack of chemical evolution in QSOs since $z \sim 6$
and suggests a formation redshift of the SN Ia progenitors of $z
\gtrsim 10$.  Extremely rapid enrichment on a time scale of just a few
hundred Myr must have occurred in these QSOs, much faster than the
slow enrichment time scales of spiral galaxies.  Nonetheless, also
here evolutionary differences are becoming apparent: Some of the $z
\sim 6$ QSOs are less evolved, {\em not}\ showing strong Fe\,{\sc ii}
emission lines (\cite[Iwamuro et al.\ 2004]{Iwamuro04}) and hence no
significant SN Ia contributions yet.  

\subsection{Galaxies at redshift $\ge 7$}

Moving to even higher redshifts, the analysis of spectral energy
distributions obtained from near-infrared imaging data of galaxies at
redshifts of 7 (age of the Universe $\sim 770$ Myr) and 8 ($\sim 640$ Myr)
revealed median ages of $\sim 200$ Myr for their stellar populations,
but even younger ages are not excluded (\cite[Finkelstein et al.\
2010]{Finkelstein10}).  The typical stellar masses of the galaxies at
$z \sim 7$ are $< 10^9$ M$_{\odot}$; they may be as low as only $10^7$
M$_{\odot}$ at $z \sim 8$.  While some galaxies are consistent with
having no internal dust extinction, the median value is $A_V \sim 0.3$
mag.  \cite[Finkelstein et al.\ (2010)]{Finkelstein10} estimate that
some $10^6$ M$_{\odot}$ of dust may have been produced in these
galaxies through massive (12 -- 35 M$_{\odot}$) stars that turned into
SNe II, a process expected to take less than 20 Myr (\cite[Todini \&
Ferrara 2001]{Todini01}).

\cite[Finkelstein et al.\ (2010)]{Finkelstein10} derive metallicities
of $0.005\,Z_{\odot}$ for the majority of their galaxies, while some
may have somewhat higher values ($0.02\,Z_{\odot}$).  Inferring
maximal stellar masses of a few times $10^9$  M$_{\odot}$ for the
$z\sim7$ and 8 galaxies, the authors emphasize that these masses are
still considerably lower than suggested for $L_\star$ counterparts at
$z < 6$.  These distant, luminous galaxies whose colors resemble those
of local, metal-poor star-bursting dwarf galaxies trace the earliest
times of high-redshift galaxy formation accessible to us with the
current instrumentation.

\section{Global evolutionary trends}

Global evolutionary trends in galaxy evolution over cosmic times
(e.g., \cite[Madau et al.\ 1996]{Madau96}) are best inferred from
individual or combined galaxy surveys covering a wide redshift range.
\cite[Marchesini et al.\ (2009)]{Marchesini09} derive the redshift
evolution of the global stellar mass density for galaxies with stellar
masses in the range of $10^8 <$ M$_\star$/M$_{\odot} < 10^{13}$.
Galaxies with lower stellar masses do not appear to contribute
significantly to the mass density budget.  Marchesini et al.\ find
that approximately 45\% of the present-day stellar mass was generated
from $3 > z > 1$ (within 3.6 Gyr). {From} $z \sim 1$ to the present,
i.e., during the last $\sim 7.5$ Gyr, the remaining 50\% were
produced.  

\subsection{Downsizing}

Considering cosmic star formation histories, there is compelling
evidence for ``downsizing'' (\cite[Cowie et al.\ 1996]{Cowie96}): The
stars in more massive galaxies usually formed at earlier cosmic epochs
and over a shorter time period.  For a pedagogical illustration, see
Fig.\ 9 in \cite[Thomas et al.\ (2010)]{Thomas10}.  High-redshift
galaxies tend to have star formation rates higher than those found in
the local Universe.  Also, these galaxies are typically more massive
than low-redshift star-forming galaxies (\cite[Seymour et al.\
2008]{Seymour08}).  

\cite[Juneau et al.\ (2005)]{Juneau05} show that the star formation
rate density depends strongly on the stellar mass of galaxies.  For
massive galaxies with M$_\star > 10^{10.8}$ M$_{\odot}$ the star
formation rate density was about six times higher at $z = 2$ than at
the present, remaining approximately constant since $z = 1$.  The star
formation rate density in their ``intermediate-mass'' bin, $10^{10.2}
\le$ M$_\star$/M$_{\odot} \le 10^{10.8}$ M$_{\odot}$, peaks at $z \sim
1.5$, whereas since $z < 1$ most of the activity was in lower-mass
galaxies (\cite[Juneau et al.\ 2005]{Juneau05}).  

For quasar host galaxies observed with the Herschel satellite,
\cite[Serjeant et al.\ (2010)]{Serjeant10} find that high-luminosity
quasars have their peak contribution to the star formation density at
$z \sim 3$, while the maximum contribution of low-luminosity quasars
peaks between $1 < z < 2$.  The authors suggest that this indicates a
decrease in both the rate of major mergers and in the gas available
for star formation and black hole accretion.   

\subsection{Mass-metallicity relations}

From an analysis of $\sim 160,000$ galaxies in the local Universe
observed by the Sloan Digital Sky Survey (SDSS), \cite[Gallazzi et
al.\ (2008)]{Gallazzi08} conclude that approximately 40\% of the total
amount of metals contained in stars is located in bulge-dominated
galaxies with predominantly old populations. Disk-dominated galaxies
contain $< 25$\%, while both types of galaxies contribute similarly to
the total stellar mass density.

\cite[Panter et al.\ (2008)]{Panter08} derived the
mass-fraction-weighted galaxian metallicity as a function of
present-day stellar mass based on the analysis of $> 300,000$ galaxies
from the SDSS.  Panter et al.\ find a flat relation with little
scatter ($\sim 0.15$ dex) around $\sim 1.1\,Z_{\odot}$ for galaxies
with masses $\gtrsim 10^{10.5}$ M$_{\odot}$.  Below $\sim 10^{10}$
M$_{\odot}$ there is a clear trend of decreasing metallicity of about
0.5 dex per dex decline in mass, although the dispersion is relatively
large ($\sim 0.5$ dex).  When considering only (the very few) galaxies
in which more than 50\% of the light comes from populations younger
than 500 Myr, only systems with stellar masses $< 10^{10}$ M$_{\odot}$
contribute significantly -- another indication of downsizing.  

\subsection{Environmental trends}

Galaxies in high-density regions are generally found to have higher
metallicities than those in low-density regions (see, e.g.,
\cite[Panter et al.\ 2008]{Panter08}).  \cite[Sheth et al.\
(2006)]{Sheth06} find that galaxies with above-average star formation
rates and high metallicities at high redshifts are situated mainly in
galaxy clusters in the present-day Universe.  Interestingly,
\cite[Poggianti et al.\ (2010)]{Poggianti10} infer that high-redshift
clusters were denser environments with respect to both galaxy number
and mass than contemporary clusters, which might have fostered the
intense star-formation activity and growth in their most massive
galaxies.  The clustering strength of star-forming galaxies decreases
with decreasing redshift (\cite[Hartley et al.\ 2010]{Hartley10}).
Galaxies that are passively evolving at the present time (typically
galaxies with halo masses $>10^{13}$ M$_{\odot}$) are twice as
strongly clustered than present-day star-forming galaxies (which are
typically at least a factor of 10 less massive).   

\cite[Sheth et al.\ (2006)]{Sheth06} point out that at lower
redshifts, star formation (either in terms of mass or fraction) is
anticorrelated with environment such that dense environments have
shown lower star formation rates than low-density regions during the
past $\sim 5$ Gyr.  Interestingly, in the present-day Universe
star-forming galaxies appear to evolve fairly independently of their
environment, with intrinsic properties playing a determining role
(\cite[Balogh et al.\ 2004]{Balogh04}; \cite[Poggianti et al.\
2008]{Poggianti08}).  Overall, environment seems to have had little
influence on the cosmic star formation history since $z < 1$
(\cite[Cooper et al.\ 2008]{Cooper08}). 

The famous morphology-density relation in clusters and groups
(\cite[Oemler 1974]{Oemler74}; \cite[Dressler 1980]{Dressler80};
\cite[Postman \& Geller 1984]{Postman84}) describes the increase in
the fraction of ellipticals with galaxy density, while the fraction of
spirals declines.  Similarly, there is a pronounced correlation of
morphological types with cluster-centric radius (\cite[Whitmore et
al.\ 1993]{Whitmore93}) such that in the innermost regions of a
cluster the fraction of ellipticals shows a strong increase, while the
fraction of spirals drops sharply.  The S0 fraction rises less steeply
with decreasing radius and also drops in the innermost regions.  These
relations are interpreted as suggestive of the growth of ellipticals
(and S0s) in high-density regions at the expense of spirals.
\cite[Goto et al.\ (2003)]{Goto03} find three different regimes
depending on galaxy density:  For densities $< 1$ Mpc$^{-2}$, the
morphology-density and morphology-radius relations become rather weak.
For 1 -- 6  Mpc$^{-2}$, the fraction of late-type disks decreases with
cluster-centric radius, while early-type spiral and S0 fractions
increase.  For the highest-density clusters with $> 6$  Mpc$^{-2}$,
also these intermediate-type fractions decrease with cluster-centric
radius, while the early-type fractions increase.  

For dwarf galaxies, similar global morphology-density and
morphology-radius trends are observed in the sense that gas-deficient
early-type dwarfs tend to concentrate within $\sim 300$ kpc around
massive galaxies in groups or are predominantly found in the inner
regions of clusters, while gas-rich late-type dwarfs are found in the
outskirts and in the field (e.g., \cite[Grebel 2000]{Grebel00};
\cite[Karachentsev et al.\ 2002a]{Karachentsev02a},
\cite[2003a]{Karachentsev03a}; \cite[Lisker et al.\ 2007]{Lisker07}).
The poorest low-density groups are dominated by late-type galaxies
both among their massive and their dwarf members (\cite[Karachentsev
et al.\ 2003a]{Karachentsev03a}, \cite[2003b]{Karachentsev03b}), while
in richer, more compact groups the early-type fractions increase also
among the dwarfs, as does the morphological segregation (see, e.g.,
\cite[Karachentsev et al.\ 2002a]{Karachentsev02a},
\cite[2002b]{Karachentsev02b}).  Various physical mechanisms including
ram pressure and tidal stripping are discussed to explain apparent
evolutionary connections of dwarf galaxies with environment (e.g.,
\cite[van den Bergh 1994]{vandenBergh94}; \cite[Vollmer et al.\
2001]{Vollmer01}; \cite[Grebel et al.\ 2003]{GreGalHar03}; \cite[Dong
et al.\ 2003]{Dong03}; \cite[Hensler et al.\ 2004]{Hensler04};
\cite[Kravtsov et al.\ 2004]{Kravtsov04}; \cite[Mieske et al.\
2004]{Mieske04}; \cite[Lisker et al.\ 2006]{Lisker06a}; \cite[Mayer et
al.\ 2006]{Mayer06}; \cite[D'Onghia et al.\ 2009]{DOnghia09}).

\section{Massive early-type galaxies}

Analyzing a volume-limited sample of $> 14,000$ early-type galaxies
from the SDSS, \cite[Clemens et al.\ (2009a)]{Clemens09a} find that
their ages, metallicities, and $\alpha$-element enhancement increase
with their mass (using velocity dispersion, $\sigma$, as indicator).
For galaxies with $\sigma > 180$ km~s$^{-1}$, the mean age decreases
with decreasing galactocentric radius, while the metallicity
increases.  Clemens et al.\ suggest that the massive early-type
galaxies were assembled at $z \lesssim 3.5$, merging with low-mass
halos that began to form at $z \sim 10$.  These subhalos contributed
older, metal-poor stars that are still distributed over large radii.
Gas-rich mergers, very frequent at early times, contributed fuel for
intense star formation in the central regions of the galaxies, while
mergers at later times were increasingly gas-poor or dry.  Clemens et
al.\ find these radial age and metallicity gradients in early-type
galaxies regardless of environment, although massive ellipticals in
clusters are on average $\sim 2$ Gyr older than those in the field,
supporting the trends expected for downsizing.  

\subsection{Environment and rejuvenation}

{From} an analysis of Spitzer Space Telescope (SST) data of 50
early-type galaxies in the Coma cluster, \cite[Clemens et al.\
(2009b)]{Clemens09b} find that while the majority is passive, some
$\sim 30$~\% of the galaxies are either younger than 10
Gyr or were rejuvenated in the last few Gyr.

Combining near-UV photometry from the Galaxy Evolution Explorer
(GALEX) satellite with SDSS data of a volume-limited sample of 839
luminous early-type galaxies, \cite[Schawinski et al.\
(2007)]{Schawinski07} conclude that $\sim 30$~\% of these objects show
evidence of recent ($< 1$ Gyr) star formation ($\sim 29$~\%
ellipticals, $\sim 39$~\% lenticulars).  Moreover, they show that that
low-density environments contain $\sim 25$~\% more UV-bright
early-type galaxies. 

\cite[Thomas et al.\ (2010)]{Thomas10} analyze low-redshift $> 3000$
early-type galaxies from the SDSS and infer that intermediate-mass and
low-mass galaxies show evidence for a secondary peak of more recent
star formation around $\sim 2.5$ Gyr ago.  They find that the fraction
of these rejuvenated galaxies becomes larger with decreasing galaxy
mass and with decreasing environmental density, reaching up to 45~\%
at low masses and low densities.  Thomas et al.\ conclude that the
impact of environment increases with decreasing galaxy mass via
mergers and interactions and has done so since $z \sim 0.2$.

\subsection{E+A galaxies}

An interesting class of rejuvenated early-type galaxies are the
so-called ``E+A'' galaxies, ellipticals that show the typical K-star
spectra with Mg, Ca, and Fe absorption lines as well as strong Balmer
lines akin to A-stars (\cite[Dressler \& Gunn 1983]{Dressler83}),
indicating that in addition to the usual passive evolution, they
experienced star formation within the last Gyr.  The absence of
[O\,{\sc ii}] and H$\alpha$ emission lines shows that there is no
ongoing star formation.  These post-starburst galaxies are observed
both in clusters and in the field.  

SDSS studies support suggestions that the E+A phenomenon is created by
interactions and/or mergers. About 30~\% of the E+A galaxies show
disturbed morphologies or tidal tails (\cite[Goto 2005]{Goto05}).  The
analysis of 660 E+A galaxies revealed that these objects have a 54~\%
higher probability of having close companion galaxies than normal
galaxies ($\sim 8$~\% vs.\ $\sim 5$~\%; \cite[Yamauchi et al.\
2008]{Yamauchi08}).

\section{Spirals and irregulars}

\subsection{Stellar halos of disk galaxies}

Spiral galaxies usually have extended, low-density stellar halos
(\cite[Zibetti et al.\ 2004]{Zibetti04}) whose density decreases with
$\sim r^{-3}$.  In the Milky Way the stellar halo consists of old,
metal-poor stars and globular clusters on eccentric prograde or
retrograde orbits (see \cite[Freeman \& Bland-Hawthorn
2002]{Freeman02} and \cite[Helmi 2008]{Helmi08} for recent reviews of
the Galactic halo).  As many of half of the field stars in the halo
may have originated in disrupted globular clusters (\cite[Martell \&
Grebel 2010]{Martell10}; see \cite[Odenkirchen et al.\
2001a]{Odenkirchen01a} for an example).  

$\Lambda$CDM simulations suggest that in galaxies with few recent
mergers the fraction of halo stars formed {\em in situ} amounts to
20~\% to 50~\% (\cite[Zolotov et al.\ 2009]{Zolotov09}). \cite[Johnston
et al.\ (2008)]{Johnston08} propose that halos dominated by very early
accretion show higher [$\alpha$/Fe] ratios, whereas those that
accreted mainly high-luminosity satellites should exhibit higher
[Fe/H].  

The detection of substructure (e.g., \cite[Newberg et al.\
2002]{Newberg02}; \cite[Yanny et al.\ 2003]{Yanny03}; \cite[Bell et
al.\ 2008]{Bell08}) as well as chemical and kinematic signatures
(\cite[Carollo et al.\ 2007]{Carollo07}; \cite[Geisler et al.\
2007]{Geisler07}) support the scenario that part of the Galactic halo
was accreted.  The individual stellar element abundance patterns
suggest that such accretion may have mainly occurred at very early
times, since the [$\alpha$/Fe] ratios of metal-poor halo stars match
the ones found in similar stars in the Galactic dwarf spheroidal and
irregular companions (e.g., \cite[Koch et al.\ 2008a]{Koch08a}).
Prominent morphological evidence of ongoing dwarf galaxy accretion has
been found not only in the Milky Way (e.g., \cite[Ibata et al.\
1994]{Ibata94}), but also around other spiral galaxies (e.g.,
\cite[Ibata et al.\ 2001]{Ibata01}; \cite[Zucker et al.\
2004]{Zucker94}; \cite[Mart\'{i}nez-Delgado et al.\
2008]{Martinez08}).  

\subsection{Bulges of disk galaxies and formation scenarios}

Early-type spirals show prominent bulges, which become less pronounced
and ultimately vanish in late-type spirals and irregulars.  Classical
bulges (found in early-type to Sbc spirals) resemble elliptical
galaxies in their properties, are dominated by old, mainly metal-rich
stars with a large metallicity spread, show hot stellar kinematics,
and follow a de Vaucouleurs surface brightness profile just like
typical elliptical galaxies.  Pseudobulges (in disk galaxies later
than Sbc) resemble disk galaxies, have similar exponential profiles,
are rotation-dominated, and may contain a nuclear bar, ring, or
spiral.  They are believed to form from disk material via secular
evolution (\cite[Kormendy \& Kennicutt 2004]{Kormendy04}).  A recent
analysis combining data from SST and GALEX found that all bulges show
some amount of ongoing star formation, regardless of their type
(\cite[Fisher et al.\ 2009]{Fisher09}), with small bulges having
formed 10 to 30~\% of their mass in the past 1 to 2 Gyr (\cite[Thomas
\& Davies 2006]{Thomas06}).  Extracting a sample of $>3000$ nearby
edge-on disk galaxies from the SDSS, \cite[Kautsch et al.\
(2006)]{Kautsch06} showed that approximately 30~\% of the edge-one
galaxies are bulge-less disks. 

\cite[Noguchi (1999)]{Noguchi99} suggested that massive clumps forming
at early times in galactic disks move towards the galactic center due
to dynamical friction, merge, and form the galactic bulge.  This
scenario leads to the observed trend of increased bulge-to-disk ratios
with increased total galactic masses.  \cite[van den Bergh
(2002)]{vandenBergh02} noted that while most of the galaxies observed
in the Hubble Deep Fields at $z < 1$ have disk-like morphologies, most
galaxies at $z > 2$ look clumpy or chaotic.  Analyzing such ``clump
clusters'' and ``chain galaxies'' in the Hubble Ultra Deep Field,
\cite[Elmegreen et al.\ (2009)]{Elmegreen09} find that the masses of
the star-forming clumps are of the order to $10^7$ to $10^8$
M$_{\odot}$.  \cite[Bournaud et al.\ (2007)]{Bournaud07} argue that
clusters of such massive, kpc-sized clumps can form bulges in less
than 1 Gyr, while the system as a whole evolves from a violently
unstable disk into a regular spiral with an exponential or double
exponential disk profile on a similarly rapid time scale.  While the
coalescence of these clumps resembles a major merger with respect to
orbital mixing, the resulting bulge has no specific dark-matter
component, which distinguishes it from bulges formed via galaxy
mergers (\cite[Elmegreen et al.\ 2008]{Elmegreen08}).  

\subsection{Disks and long-term evolutionary trends for disk galaxies}

Disks are the primary sites of present-day star formation in spiral 
galaxies, and it seems likely that they have continued to form stars 
for a Hubble time.  Disks show ordered rotation, and their stars move 
around the galactic center on near-circular orbits.  The rotational
velocities greatly exceed the velocity dispersion by factors of 20
or more.   

Gas-deficient early-type disk galaxies show little activity at the
present time, while gas-rich late-type disks experience wide-spread,
active star formation.  Star formation occurs mainly in the midplane
of the thin disks, in particular along spiral arms, where recent
events are impressively traced by giant H\,{\sc ii} regions.  Spiral
density waves may induce star formation (see
\cite[Mart{\'{\i}}nez-Garc{\'{\i}}a et al.\ 2009]{Martinez-Garcia09};
and references therein), although it has been suggested that this
mechanism may contribute less than 50~\% to the overall star formation
rate (\cite[Elmegreen \& Elmegreen 1986]{Elmegreen86}).  

In the Milky Way, the star formation in the disk was not constant, but
shows extended episodes of increased and reduced activity (e.g.,
\cite[Rocha-Pinto et al.\ 2000]{Rocha-Pinto00}), a radial metallicity
gradient, a G-dwarf problem, and a large metallicity scatter at all
ages (\cite[Nordstr\"om et al.\ 2004]{Nordstroem04}).  The thin disk
is embedded in a lower-density, kinematically hotter stellar
population consisting of older, more metal-poor stars -- the thick
disk (\cite[Gilmore \& Reid 1983]{Gilmore83}; \cite[Bensby et al.\
2005]{Bensby05}). The chemical similarity of Galactic bulge
and thick disk stars might suggest that the Milky Way does not have a
classical bulge (\cite[Mel\'endez et al.\ 2008]{Melendez08}).  

\cite[Dalcanton \& Bernstein (2002)]{Dalcanton02} showed that thick
disks are ubiquitous also in bulge-less late-type disk galaxies, which
indicates that their formation is a universal property of disk
formation independent from the formation of a bulge.  A variety of
mechanisms for the formation of thick disks has been proposed,
including formation from accreted satellites, gas-rich mergers,
heating of an early thin disk by mergers, heating via star formation
processes, and radial migration (e.g., \cite[Wyse et al.\
2006]{Wyse06}; \cite[Brook et al.\ 2004]{Brook04}; \cite[Kroupa
2002]{Kroupa02}; \cite[Bournaud et al.\ 2009]{Bournaud09}; \cite[Ro{\v
s}kar et al.\ 2008]{Roskar08}).  \cite[Sales et al.\ (2009)]{Sales09}
suggest that the eccentricity distribution of thick disk stars may
permit one to distinguish between these scenarios. 


In cosmological simulations disk galaxies may form, for example, via
major, wet mergers (see, e.g., \cite[Barnes 2002]{Barnes02};
\cite[Governato et al.\ 2009]{Governato09}) or without 
mergers via inside-out and vertical collapse in a growing dark matter
halo (\cite[Samland \& Gerhard 2003]{Samland03}).

As noted by \cite[van den Bergh (2002)]{vandenBergh02} based on an
analysis of the Hubble Deep Fields, roughly one third of the objects
at $z>2$ seem to be experiencing mergers.  He suggests that from $1 <
z < 2$ a transition from merger-dominated to disk-dominated star
formation occurred.  Moreover, he finds that at $z > 0.5$, there are
fewer and fewer barred spirals.  While early-type galaxies assume
their customary morphologies relatively early on, 46~\% of the spirals
at $0.6 < z < 0.8$ are still peculiar, and with higher redshift, the
spiral arm patterns become increasingly chaotic.  Also within the
class of spiral galaxies there are trends:  Only $\sim 5$~\% of the Sa
and Sab galaxies are peculiar at $z \sim 0.7$, while almost 70~\% of
the Sbc and Sc types are still peculiar.  

\cite[Elmegreen et al.\ (2007)]{Elmegreen07} suggest that the
formation epoch of clumpy disk galaxies may extend up to $z \sim 5$.
The ones experiencing major mergers may form red spheroidals at $2
\lesssim z \lesssim 3$, whereas the others evolve into spirals.
Elmegreen et al.\ propose that the the star formation activity in
clumpy disks is caused by gravitational collapse of portions of the
disk gas without requiring an external trigger.

Regarding environment, \cite[Poggianti et al.\ (2009)]{Poggianti09}
find that the fraction of ellipticals remains essentially constant
below $z = 1$, while the spiral and S0 fractions continue to evolve,
showing the most pronounced evolution in low-mass galaxy clusters.
They attribute this to secular evolution and to environmental
mechanisms that are more effective in low-mass environments.  At low
redshifts, the declining spiral fraction with density is driven by
late-type spirals (Sc and later; \cite[Poggianti et al.\
2008]{Poggianti08}).  

\subsection{Irregulars}

Irregular galaxies are gas-rich, low-mass, metal-poor galaxies without
spiral density waves, which show recent or ongoing star formation that
appears to have extended over a Hubble time (\cite[Hunter
1997]{Hunter97}).  Many studies found the H\,{\sc i} gas to be
considerably more extended than the stellar component in irregulars
(e.g., \cite[Young \& Lo]{Young97}), but more recent, deep optical
surveys show that the optical extent of at least some of these
galaxies has been underestimated (e.g., \cite[Kniazev et al.\
2009]{Kniazev09}).  All nearby irregulars and dwarf irregulars have
been found to contain old populations, although their fractions differ
(\cite[Grebel \& Gallagher 2004]{GreGal04}).  The old populations tend
to be more extended than the more recent star formation (e.g.,
\cite[Minniti \& Zijlstra 1996]{Minniti96}; \cite[Kniazev et al.\
2009]{Kniazev09}) and show a more regular distribution (e.g.,
\cite[Zaritsky et al.\ 2000]{Zaritsky00}; \cite[van der Marel
2001]{vanderMarel01}).

Irregulars are usually found in the outskirts of groups and clusters
or in the field, thus interactions with other galaxies are likely to
be rare.   Their star formation appears to be largely governed by
internal processes and seems to be stochastic.  Rather than
experiencing brief, intense starbursts, irregulars typically show
extended episodes of star formation interrupted by short quiescent
periods -- so-called gasping star formation (e.g., \cite[Cignoni \&
Tosi 2010]{Cignoni10}).  The long-term star formation amplitude
variations amount to factors of 2 to 3 (\cite[Tosi et al.\
1991]{Tosi91}).  For a review of irregulars and dwarf irregulars in
the Local Group, for which we have the most detailed data to date, see
\cite[Grebel (2004)]{Grebel04}.  

While the more massive irregulars are rotationally supported and show
solid-body rotation, low-mass dwarf irregulars are dominated by random
motions.  Star formation ceases at lower gas density thresholds than
in spirals (e.g., \cite[Parodi \& Binggeli 2003]{Parodi03}), and the
global gas density of the highly porous interstellar medium has been
found to lie below the Toomre criterion for star formation (\cite[van
Zee et al.\ 1997]{vanZee97}).  Turbulence may create local densities
exceeding the star formation threshold (e.g., \cite[Stanimirovic et
al.\ 1999]{Stanimirovic99}). Low-mass dwarf irregulars without
measurable rotation show less centrally concentrated star formation
and have lower star formation rates (\cite[Roye \& Hunter
2000]{Roye00}; \cite[Parodi \& Binggeli 2003]{Parodi03}).  In most
irregulars, star formation occurs within the galaxies' Holmberg radius
and within three disk scale lengths (\cite[Hunter \& Elmegreen
2004]{Hunter04}).  

In contrast, in blue compact dwarf (BCD) galaxies the highest star
formation rates are found, and star formation occurs mainly in the
central regions (Hunter \& Elmegreen).  (We do not discuss BCDs and
other gas-rich dwarfs in more detail here. For an overview of
different dwarf types and their properties, see \cite[Grebel
(2003)]{Grebel03}).    

Once believed to be chemically homogeneous, there is now evidence of
metallicity variations at a given age in several irregulars (e.g.,
\cite[Kniazev et al.\ 2005]{Kniazev05}; \cite[Glatt et al.\
2008]{Glatt08}).  This suggests that local processes dominate the
enrichment and that mixing is not very efficient.  Irregulars follow a
fairly well-defined metallicity-luminosity relation, which however is
offset from that of early-type dwarfs covering the same luminosity
range.  Surprisingly, the offset is such that the continuously
star-forming irregulars and dwarf irregulars have lower metallicities
at a given luminosity than the inactive early-type dwarfs (e.g.,
\cite[Richer et al.\ 1998]{Richer98}), a discrepancy that holds even
when comparing stellar populations of the same age (\cite[Grebel et
al.\ 2003]{GreGalHar03}).  Taken at face value, this may imply that
the enrichment of irregulars was less efficient and slower than that
of early-type dwarfs.  BCDs and in particular extremely
metal-deficient galaxies continue this trend and appear to be too
luminous for their present-day, low metallicities even when compared
to normal irregulars (\cite[Kunth \& \"Ostlin 2000]{Kunth00};
\cite[Kniazev et al.\ 2003]{Kniazev03}).    

\subsection{Star formation ``demographics''}

\cite[Lee et al. (2007)]{Lee07} investigate the star formation
``demographics'' of star-forming galaxies out to 11 Mpc combining
H$\alpha$ and GALEX UV fluxes.  Their sample includes spirals,
irregulars, and BCDs.  Lee et al.\ identify three different star
formation regimes: 

1. Galaxies with maximum rotational velocities $V_{max} > 120$
km~s$^{-1}$, total B-band magnitudes of $M_B \lesssim -19$, and
stellar masses $\gtrsim 10^{10}$ M$_{\odot}$ are mainly
bulge-dominated galaxies with relatively low specific star formation
rates and increased scatter in these rates.  Also the mass-metallicity
relation changes its slope in this regime (\cite[Panter et al.\
2007]{Panter07}), and supernova ejecta can be retained (\cite[Dekel \&
Woo 2003]{Dekel03}).  \cite[Bothwell et al.\ (2009)]{Bothwell09} find
that the H\,{\sc i} content of these massive galaxies decreases faster
than their star formation rates, leading to shorter H\,{\sc i}
consumption time scales and making the lack of gas a plausible reason
for the observed quenching of star formation activity. 

2. Galaxies with $\sim 120$ km~s$^{-1} > V_{max} > 50$ km~s$^{-1}$ and
$-19 < M_B < -15$ comprise mainly late-type spirals and massive
irregulars.  \cite[Lee at al.\ (2007)]{Lee07} suggest that spiral
structure acts as an important regulatory factor for star formation.
They find that the galaxies in this intermediate-mass regime exhibit a
comparatively tight, constant relation between star formation rate and
luminosity (or rotational velocity).  The star formation rates show
fluctuations of 2 to 3, and the current star formation activity is
about half of its average value in the past.  \cite[Bothwell et al.\
(2009)]{Bothwell09} argue that the galaxies in this regime evolve
secularly. They show that the star formation rates decrease
with the galaxies' H\,{\sc i} mass, and that the H\,{\sc i}
consumption time scales increase with decreasing luminosity.  

3. Below $V_{max} = 50$ km~s$^{-1}$ and $M_B > -15$, dwarf galaxies,
particularly irregulars, dominate.  At these low masses, the star
formation rates exhibit much more variability ranging from
significantly higher (e.g., in BCDs) to significantly lower (e.g., in
so-called transition-type dwarfs with properties in between dwarf
irregulars and dwarf spheroidal galaxies, see \cite[Grebel et al.\
2003]{GreGalHar03}) star formation activity than in the higher-mass
regimes. Overall, there is a general trend towards lower star
formation rates.  Stochastic intrinsic processes, feedback, and the
ability to retain gas play an important role here.  \cite[Bothwell et
al.\ (2009)]{Bothwell09} find that for many of the galaxies in the
low-mass regime the H\,{\sc i} consumption time scale exceeds a Hubble
time (in good agreement with the results of \cite[Hunter
1997]{Hunter97}).  

Bothwell et al.\  show that the H\,{\sc i} consumption time scales
have a minimum duration of more than 100 Myr.  They argue that this
minimum duration corresponds to the gas mass divided by the minimum
gas assembly time, i.e., the free-fall collapse time.   

\section{Early-type dwarfs}

We end this review by summarizing the star formation properties of
early-type dwarfs, most notably of dwarf ellipticals (dEs) and dwarf
spheroidals (dSphs; see \cite[Grebel 2003]{Grebel03}; \cite[Grebel et
al.\ 2003]{GreGalHar03}).  Typically located in high-density regions
such as the immediate surroundings of massive galaxies or in galaxy
clusters, this dense environment may have affected the evolution of
these now gas-deficient galaxies (Section 3.3).  Structural and
kinematic studies suggest that early-type dwarfs are strongly
dark-matter dominated (e.g., \cite[Odenkirchen et al.\
2001b]{Odenkirchen01b}; \cite[Klessen et al.\ 2003]{Klessen03};
\cite[Wilkinson et al.\ 2004]{Wilkinson04}; \cite[Koch et al.\
2007a]{Koch07a}, \cite[2007b]{Koch07b}; \cite[Walker et al.\
2007]{Walker07}; \cite[Gilmore et al.\ 2007]{Gilmore07}; \cite[Wolf et
al.\ 2010]{Wolf10}), and there are even indications of a constant
dark-matter halo surface density from spirals to dwarfs (\cite[Donato
et al.\ 2009]{Donato09}). However, it is not yet clear whether the
apparently constant total mass regardless of a dwarf's baryonic
luminosity is universal (e.g., \cite[Ad\'en et al.\ 2009]{Aden09}).   

All of dEs and dSphs studied in detail so far reveal varying fractions
of old populations (\cite[Grebel \& Gallagher 2004]{Grebel04};
\cite[Da Costa et al.\ 2010]{DaCosta10}) that become dominant at low
galactic masses, while intermediate-age populations ($> 1$ Gyr) are
prominent at higher galactic masses.  Still, no two dwarfs share the
same evolutionary history or detailed abundance properties
(\cite[Grebel 1997]{Grebel97}).  Population gradients are found in
many early-type dwarfs.  Where present, younger and/or more metal-rich
populations tend to be more centrally concentrated (e.g.,
\cite[Harbeck et al.\ 2001]{Harbeck01}; \cite[Lisker et al.\
2006]{Lisker06b}; \cite[Crnojevi\'c et al.\ 2010]{Crnojevic10}).   

The Galactic dSphs reach solar [$\alpha$/Fe] ratios at much lower
[Fe/H] than typical Galactic halo stars, which suggests low star
formation rates, the loss of metals and supernova ejecta, and/or a
larger contribution from SNe Ia (e.g., \cite[Shetrone et al.\
2001]{Shetrone01}).  Abundance spreads of 1 dex in [Fe/H] and more are
common.  The scatter in $\alpha$ element abundance ratios at a given
metallicity underlines the inhomogeneous, localized enrichment in the
early-type dwarfs, another characteristic expected of slow, stochastic
star formation and low star formation efficiencies (\cite[Koch et al.\
2008a]{Koch2008a}, \cite[2008b]{Koch2008b}; \cite[Marcolini et al.\
2008]{Marcolini08}).

\end{document}